\documentclass[amsmath,amssymb,aps,floatfix,letterpaper,prl,superscriptaddress,twocolumn,showpacs]{revtex4}

\usepackage{bm}
\usepackage{graphicx}
\usepackage{amsmath}
\usepackage{amssymb}
\usepackage{dcolumn}
\usepackage[caption=false]{subfig}  
\usepackage{afterpage}
\usepackage{color}

\begin{document}

\title{Heat conduction across molecular junctions between nanoparticles}

\date{\today}

\author{Samy Merabia}
\email{samy.merabia@univ-lyon1.fr}
\affiliation{LPMCN, Universit\'e de Lyon; UMR 5586 Universit\'e Lyon 1 et CNRS, F-69622 Villeurbanne, France}
\author{Jean-Louis Barrat}
\email{jean-louis.barrat@ujf-grenoble.fr}
\affiliation{LPMCN, Universit\'e de Lyon; UMR 5586 Universit\'e Lyon 1 et CNRS, F-69622 Villeurbanne, France}
\affiliation{LIPHY,              Universit\'e Grenoble 1 and CNRS,
               UMR 5588, Saint Martin d'H\`eres, F-38402, France}
\author{Laurent J. Lewis}
\email{laurent.lewis@umontreal.ca}
\affiliation{D\'epartement de Physique et Regroupement Qu\'eb\'ecois sur les
Mat\'eriaux de Pointe (RQMP), Universit\'e de Montr\'eal, C.P. 6128, Succursale
Centre-Ville, Montr\'eal, Qu\'ebec, Canada H3C 3J7}

\begin{abstract}

We investigate the problem of heat conduction across a molecular junction
connecting two nanoparticles, both in vacuum and in a liquid environment,
using classical molecular dynamics simulations. In vacuum, the well-known
result of a length independent conductance is recovered; its precise value,
however, is found to depend sensitively on the overlap between the
vibrational spectrum of the junction and the density of states of the
nanoparticles that act as thermal contacts. In a liquid environment, the
conductance is constant up to a crossover length, above which a standard
Fourier regime is recovered.

\end{abstract}


\maketitle

\section{Introduction}


Since the seminal work of Fermi, Pasta and Ulam \cite{fermi1955}, the problem
of heat transfer in low dimensional systems has attracted much attention
(see, e.g., Refs.\ \cite{cahill2003,dhar2008} for a review), in particular
with regards to nanoscale systems. While the problem is fundamentally
important and interesting, understanding heat transport in nanoscale systems
(including molecular junctions) is of considerable importance for
applications, e.g., in nano-electronic as well as thermo-electric devices
where, typically, a minimal thermal conductivity is
required to convert heat in electric current;
molecular junctions are also promising systems for use as thermal rectifiers
\cite{chang2006,casati2007}.

The problem has been the object of numerous theoretical studies, which may be
classified in two categories: (i) simple, toy-model one-dimensional systems
--- typically a chain --- coupled to heat baths
\cite{dhar2006,segal2003,zhou2010}; (ii) numerical simulations of realistic
systems. 
 The latter category includes, among others, molecular
dynamics studies of carbon nanotubes \cite{mingo2005}, self-assembled
monolayers \cite{luo2010}, molecular junctions \cite{mingo2006} and
polyethylene chains as well \cite{henry2008}, which have been motivated to a
large extent by experimental studies of a number of materials, including
nanowires~\cite{tighe1997,schwab2000}, carbon
nanotubes\cite{berber2000,kim2001,yu2005} and diamond
nano-rods~\cite{padgett2006}.

From a fundamental viewpoint, nanoscale heat transport is closely related to
the study of the validity  of Fourier's law at small spatial  scales. In the context of  a junction of length $L$
subject to a temperature difference $\Delta T$, the heat current $J$ is given by 
   \begin{equation}
   \label{fourier}
   J =  \kappa \frac{\Delta T}{L},
   \end{equation}
where  $\kappa$ is the thermal conductivity, which in general is a material property. In short junctions, the physical
dimensions of the systems of interest is comparable to the phonon mean free
path $\Gamma$, and one expects a transition from a ballistic regime, where
$\kappa \sim L$ when $L < \Gamma$, to
a diffusive regime when $L$ is large and temperature is high; in this case,
$\kappa \sim L^{\alpha}$ where $0 \le \alpha <1$, which may eventually lead
to the divergence of the conductivity with system size  for systems with vanishing
concentrations of impurities.

The scaling of the thermal conductivity with length is often analysed  using the
Landauer formula~\cite{landauer1970} commonly used in mesoscopic physics:
   \begin{equation}
   \label{landauer1}
   J = \int_{0}^{\infty} \frac{\bar{h}\omega}{2 \pi} \mathcal T (\omega)
   (f_L-f_R) d \omega,
   \end{equation}
where $\mathcal T(\omega)$ is the transmission coefficient of phonons, and
$f_L$ and $f_R$ are the distributions of phonons in the two heat baths
(`left' and `right') to which the system is connecting. These distributions depend
on the densities of vibrational states and on the thermal  (Bose Einstein or classical) statistics
In the limit where the temperature difference
$\Delta T$ between the two heat baths vanishes and  if the two baths have identical statistics, the Landauer formula becomes
consistent with the Fourier law:
   \begin{equation}
   \label{landauer2}
   J = \left(\int_{0}^{\infty} \frac{\bar{h}\omega}{2 \pi} \mathcal T
   (\omega) \frac{\partial f}{\partial T} d \omega \right) \Delta T,
   \end{equation}
with a conductivity increasing linearly with length: $\kappa \sim L$.

The dependence of  the thermal conductivity on the system length has been extensively studied
in nanowires and carbon nanotubes~\cite{dhar2008}. There has been
comparatively fewer studies in molecular junctions, largely because
measuring heat currents across molecular junctions is extremely difficult. The usual way to
measure heat current across a 1D object consists in connecting the object to
two electrodes and measuring the heating Joule effect due to a difference of
voltage between the electrodes. However, this gives access to the thermal
conductance and one therefore needs a model for the nanoscale junction to
extract the conductivity. Also, the thermal contacts of the junction may
contribute to the Joule effect, thus making the determination of the
conductance inaccurate. A promising but less explored method consists in
using nanoparticles (NPs) as thermal contacts.  Indeed,  NPs of identical or
different materials can be permanently linked with a
few nanometers long molecule --- e.g. an alcenes or complementary DNA strands. Such particle may then serve as thermal 
contacts, either to be inserted within a NEMS device for thermal transport, or manipulated, heated and thermally monitored 
using optical methods.

In the present article, we propose a numerical protocol~\cite{merabia2009a,merabia2009b} 
to measure the thermal conductivity of a
molecular junction permanently linking two NPs, possibly immersed in a
liquid. We examine the steady state current/temperature difference for the
junction, and we compare the conductivity in the linear regime for junctions
in vacuum and in a liquid. We unveil the crucial role of the liquid solvent
on the thermal conductivity of the junction with respect to its length. For a
junction in vacuum, i.e., ``dry'', the thermal conductivity is found to
increase linearly with the junction length $L$, in agreement with the
Landauer formula, eq.\ref{landauer2}. For a junction immersed in a liquid
solvent, or ``wet'', in contrast, we find a different behaviour: when the
junction is short, the conductivity increases approximately linearly with
$L$, as for the dry case, then saturates for longer junctions; the difference
results from the ``friction'' of the solvent, which damps the low-frequency
modes of the junction.

\section{Computational details}

\begin{figure}
\includegraphics[width=0.7\linewidth]{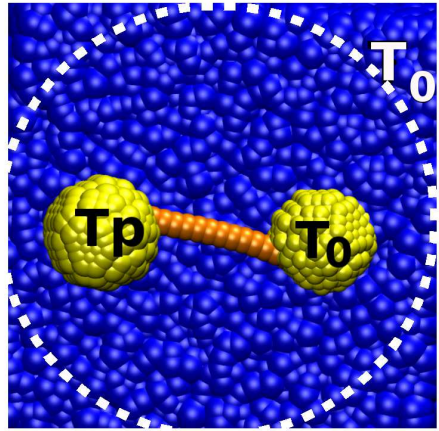}
\caption{
Snapshot of a wet junction (with $N=13$ beads) between two NPs embedded in a
Lennard Jones liquid. The region beyond the dashed line is thermostatted at
$T_0$.
}
\label{snapshot}
\end{figure}

\begin{figure}
\includegraphics[width=0.75\linewidth]{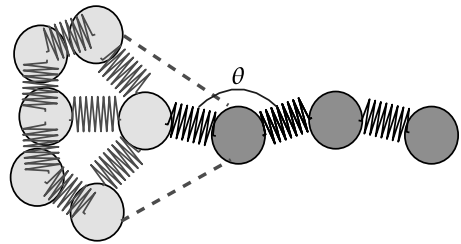}
\caption{
Schematic representation of the junction. The dark circles represent the
beads forming the junction, which interact via Hookean springs (black wavy
lines); the light circles represent NP atoms and these interact
through FENE springs. The soft springs between the NPs and the
junction are indicated by dashed lines.
}
\label{setup}
\end{figure}

We have used molecular dynamics (MD) simulations to model a molecular
junction between two NPs, embedded or not in a liquid matrix, as illustrated
in Fig.\ \ref{snapshot}. The NPs each contain $555$ atoms  cut from an FCC lattice, interacting via
the combination of a simple Lennard-Jones interaction and a nonlinear spring that forms the so called
FENE interaction \cite{kremer1990}. The stiffness and maximum extension of the nonlinear spring
are  $k_{\rm FENE}=30 \epsilon/\sigma$
and $l_{\rm max}=1.5 \sigma$, with $\sigma$ and $\epsilon$
the diameter and energy of the Lennard Jones potential (see
below). The molecular junction, whose setup is displayed in Fig. \ref{setup},
consists of $N$ beads ($N=5-25$); each bead interacts with its neighbours via
a Hookean spring of stiffness $k$, which we varied, and equilibrium length
$l_0=1.12\sigma$, exactly at the minimum of the Lennard Jones potential. The
end atoms of the junction are connected each to one atom from the two NPs so
that the equilibrium length of the junction is $L_0 = (N+1)l_0$. However, in
order to keep the junction locally perpendicular to the NP, second-neighbour
interactions are added in the form of soft springs of stiffness $k=30
\epsilon/\sigma$ and equilibrium length $l'_0 = 1.59 \sigma$. Apart from the
elastic springs, the atoms of the junction interact with their neighbours
through the bending energy term $\mathcal H = \kappa (1 + \cos \theta)$,
where $\theta$ is the angle between two
successive segments, and $\kappa$ is the bending constant which sets the  rigidity of the chain; in what
follows, we set $\kappa =1000 \epsilon$ which yields a stiff molecular
junction with a persistence length $l_p \simeq 2000 l_0$ much larger than the
equilibrium length of the junction, even for the longest junctions considered
($N=25$). Finally, each bead of the junction interacts with the atoms in the
NPs through the Lennard-Jones potential $V_{\rm LJ}(r)=4 \epsilon
((\sigma/r)^{12} - (\sigma/r)^6)$, where $\epsilon$ and $\sigma$ fix the
units of energy and length, respectively, which we set to $1$, as we also do
for the mass $m$ of all atoms. For wet junctions, the particles in the liquid
interact via Lennard-Jones potentials with the same parameters $\epsilon$ and
$\sigma$, as do the the atoms of both the junction and the NPs with the
liquid. In what follows, all quantities are expressed in Lennard-Jones units
(LJU), which are based on the elementary units  $\epsilon$, $\sigma$ and $m$.

To prepare the wet junction in an equilibrium state, the whole system is
first equilibrated at temperature $T_0=0.75 \epsilon/k_B$ under pressure
$P_0=0.0015 \epsilon/\sigma^2$ using a Nos\'e-Hoover thermostat. For the dry
junction, we thermalize the junction and the NPs at the temperature
$T_0$ and remove the total angular momentum of the system to avoid the
possibility of a permanent rotational motion. To speed up equilibration in the
latter case, we have applied a small friction force to all atoms in the
system, running over $500,000$ time steps ($dt=0.002 \sqrt{m
\sigma^2/\epsilon}$). After this initial equilibration period, for both wet
and dry junctions, one of the two NPs is heated up to constant temperature
$T_p>T_0$ using velocity rescaling while the temperature of the other NP is
maintained at $T_0$. In presence of a liquid matrix, the region at distance
$>10 \sigma$ from the surface of the NP, schematically represented in
Fig.~\ref{snapshot}, is thermostatted at $T_0$; the whole system is also
maintained at constant pressure $P_0$. Under these conditions, the
energy transferred from one NP to the other across the junction corresponds
to the amount of energy necessary to maintain the cold NP at $T_0$.

\section{Conductivity of dry versus wet junctions}

\begin{figure}
\includegraphics[width=0.9\linewidth]{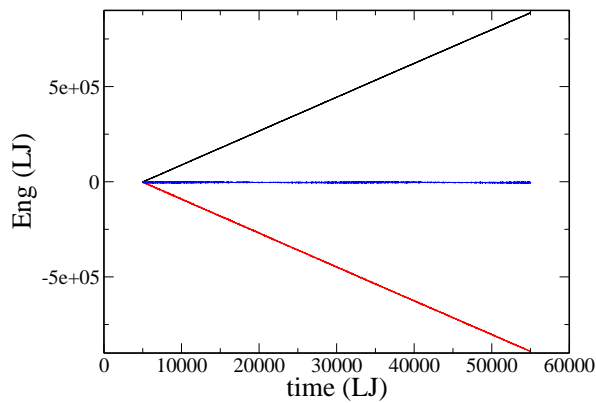}
\caption{
Heat conduction across the dry, $N=5$ junction with $k=6000$ as a function of
time: energy supplied to the hot NP (black line), energy removed from the
cold NP (red line) and sum of the two (blue line).}
\label{eng_junction_dry}
\end{figure}

Figure \ref{eng_junction_dry} displays the energy transferred across a short
($N=5$) dry junction with $k=6000$ as a function of time. The spring
stiffness is chosen here such that the characteristic pulsation (angular
frequency), $\omega=\sqrt{k/m}$, matches the density of states (DOS) of the
NPs, as will be discussed below. The junction is found to conduct energy;
Fig.~\ref{eng_junction_dry} shows that almost immediately after heating the
junction, the energy supplied to the hot NP is exactly compensated by the
energy removed from the cold NP, and the junction enters a steady state
regime: the energy transferred from one NP to the other increases linearly
with time, i.e., the power $P$ flowing through the junction is constant.

\begin{figure}
\includegraphics[width=0.9\linewidth]{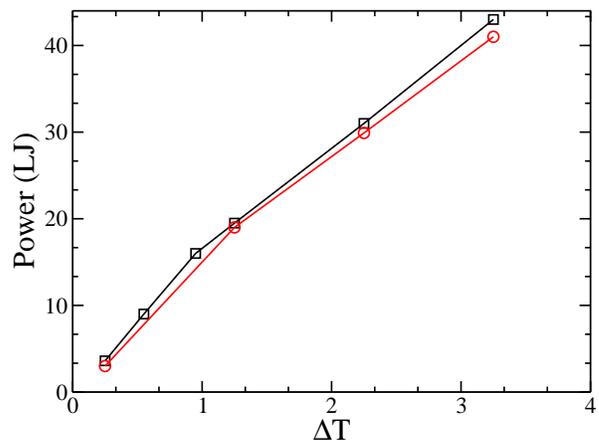}
\caption{
Power as a function of the temperature difference $\Delta T$ for a short
($N=5$) dry junction ($\circ$, black line) and wet junction ($\square$, red
line). The lines are guides to the eye.
}
\label{power_T}
\end{figure}

The corresponding results for the power as a function of the temperature
difference $\Delta T$ between the hot and cold NP are presented in
Fig.~\ref{power_T}. As expected, the power increases with $\Delta T$. A
linear behaviour,  is observed for temperature difference smaller than $1$ LJU,
which corresponds in real units to hundreds of Kelvins. This linear behaviour allows one to define 
unambiguously the conductance of the junction. With these
temperature differences, the typical power is a few tenths of nW. For larger
$\Delta T$'s, the power increases more slowly, a consequence of anharmonic
contributions. Indeed, the atoms of the NP are connected with non linear
springs which, at high temperature, may couple strongly with the springs from
the junction, making the hole system strongly anharmonic. Fig.~\ref{power_T} shows also
that the energy transfer is not affected by the presence of a liquid
surrounding the system, at least for short junctions; this suggests that the
heat flow via the liquid is negligible compared to that across the junction.

To verify this statement, we have run simulations without a junction, i.e. for 
two NPs at a surface-to-surface distance equal to the equilibrium length of
the $N=5$ junction considered above, $L_0 = 6 l_0 \simeq 7 \sigma$. In
practice, this is achieved by attaching the center of mass of each NP to a
fixed point using a stiff spring. The two NPs are immersed in the same liquid
as the NP-junction system and equilibrated, after which one of the NP is
heated up to temperature $T$ while the other is thermostatted to $T_0$ (and
so is the liquid at distance $>10 \sigma$). In this situation, the liquid
between the two NPs conducts energy due to the local temperature gradient
(which is $\sim1$ K/nm!). The conductance of the liquid is obtained by
measuring the energy necessary to thermalize the cold NP.

\begin{figure}
\includegraphics[width=0.9\linewidth]{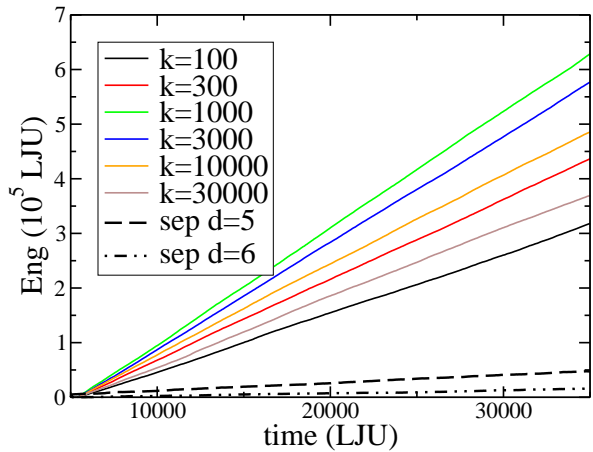}
\caption{
Energy transferred across the wet junction compared to the energy transferred
across the liquid, for a short ($N=5$) junction and for a pure liquid with a
comparable interparticle separation, respectively $d=5$ (dashed lines) and $d=6 \sigma$ (dashed-dotted lines).
 The temperature difference between the two contacts of the junction is $\Delta T=1.15$.
}
\label{wet_junction}
\end{figure}

Fig.~\ref{wet_junction} provides a comparison of the energy transferred
across short wet junctions for different values of the spring constant $k$ to
the energy transferred across the liquid alone. The range of values of $k$
has been chosen so as to cover all possible situations of overlap between the
DOS of the NPs and the vibration spectrum of the junction. In particular, for
the lowest value ($k=100$), the characteristic pulsation $\omega =\sqrt{k/m}$
is smaller than the low frequency branch of the NP DOS, and one therefore
expects weak energy transfer in this case. Indeed, the energy transferred
across the liquid is  negligibly small compared to the energy
transferred across the junction. The power flowing across the $1$ nm thick
liquid layer is approximately $1$ nW.

So far, we have considered a perfectly wetting (hydrophilic) situation. For
large contact angles (i.e., a more hydrophobic particle) --- which can be
achieved by tuning the attractive part of the NP-liquid interaction --- one
expects even smaller energy transfer capability across the liquid. Indeed, at
the interface between the NPs and the liquid, the temperature field displays
a discontinuity due to the finite conductance of the
interface\cite{barrat1999}; this discontinuity is known to increase with
hydrophobicity, as the mismatch between the two media decreases the
transmission of phonons through the interface. For our two NPs and given the
length $L$ of the junction, the effective temperature gradient in the liquid
decreases as hydrophobicity increases because the temperature jump at the
NP-liquid gets larger (i.e., the temperature of the liquid is smaller).
Hence, the effective conductance of the liquid is expected to be even lower
than in the hydrophilic case.

\section{Influence of the stiffness}

\begin{figure}
\includegraphics[width=0.9\linewidth]{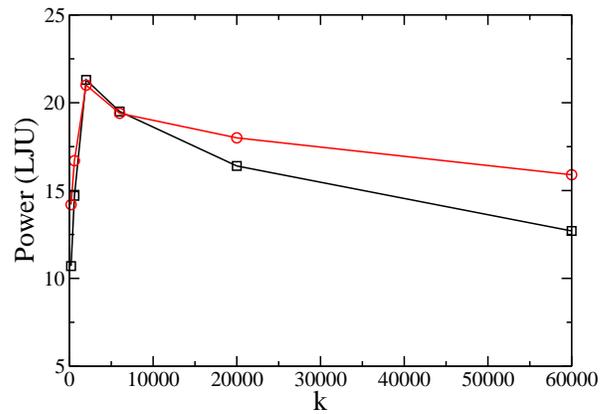}
\caption{
Power as a function of spring stiffness $k$, for wet ($\square$) and dry
($\circ$) short ($N=5$) junctions; $\Delta T=1.15$.
}
\label{power_stiffness}
\end{figure}

We now examine the variation of the conductance with the stiffness of the
springs connecting the atoms within the junction. The results are presented
in Fig.~\ref{power_stiffness} for both the wet and the dry $N=5$ junction.
The dependence on $k$ is found to be non monotonous in both cases: the power
first increases, reaches a maximum --- at the same value of $k$ ($\simeq
2000$) for both wet and dry junctions ---, then decreases, the dry junction
thus appearing to be a better conductor than the wet junction for a large
stiffness.

\begin{figure}
\includegraphics[width=0.7\linewidth,height=0.6\linewidth]{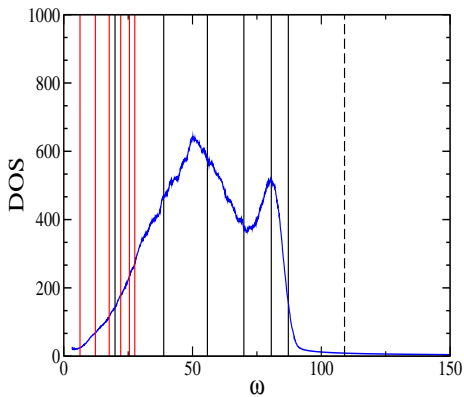}
\caption{
Vibrational spectrum of a $N=5$ junction (vertical lines) superimposed on the
NP DOS (full line) for different values of the longitudinal stiffness $k$:
$k=200$ (red lines), $2000$ (black lines) and $60000$ (dashed line in this
case, only the lowest frequency mode is shown).
}
\label{DOS_NP_junction}
\end{figure}

This particular behaviour can be explained  qualitatively in terms of the DOS
of the NPs and the vibrational spectrum of the junction, as depicted in
Fig.~\ref{DOS_NP_junction}. The DOS of the NPs has been obtained by Fourier
transforming the velocity autocorrelation function. For the junction, we have
only considered, for the sake of clarity, the spectrum of stretching
modes, whose resonance frequencies are given by $\omega_p=\sqrt{k/m} \sin(\pi
p/2N)$, $p$ being an integer between $1$ and $N-1$. Apart for these $N-1$
modes, there are $N-2$ resonance frequencies corresponding to bending which,
in the harmonic approximation, are decoupled from  longitudinal modes so that
the corresponding spectrum does not depend on $k$. Different cases may be
encountered, depending on $k$, as demonstrated in Fig.~\ref{DOS_NP_junction}.
For small values of $k$, only  a few  modes of the junction overlap with the low frequency
branch of the NP DOS, so that only the low frequency modes of the NP are
excited and a small conductance is expected, as indeed observed in
Fig.~\ref{power_stiffness}. When $k \simeq 2000$, the matching between the
stretching modes and the NP DOS is optimal, and the junction is a very good
conductor, consistent with the results of Fig.~\ref{power_stiffness}. For
large values of $k$, the normal modes of the junction have high frequencies
and there is little overlap with the NP DOS so that the energy transfer
across the junction is expected to decline.

It turns out, however, that the conductance of stiff junctions is larger than
that of soft junctions, either dry or wet (cf.\ Fig.~\ref{power_stiffness}).
This may appear counterintuitive, as there is some overlap between the
soft-junction spectrum and the NP DOS while there is none for the stiff
junction. A possible explanation is that, for large stiffness, the
high-frequency longitudinal modes of the junction excite higher harmonics of
the NPs, allowing energy to be transferred between them. As noted earlier,
Fig.~\ref{power_stiffness} also reveals that the dry junction conducts heat
better than the wet junction. The difference may be understood in terms of
the friction of the liquid, which damps the low frequency modes of the NPs,
thus reducing the efficiency of the non-linear coupling discussed above.

\begin{figure}
\includegraphics[width=0.9\linewidth]{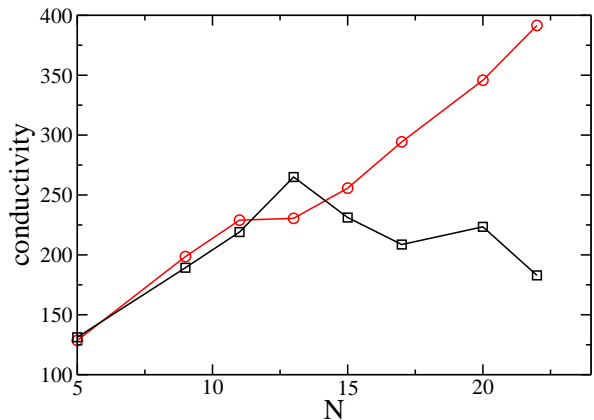}
\caption{
Conductivity of dry ($\circ$) and wet ($\square$)junctions as a function of
length, measured by the number of beads of the junction $N$. Here, $k=6000$
and $\Delta T=1.15$.}
\label{power_length}
\end{figure}

\section{Effect of length}

\begin{figure}
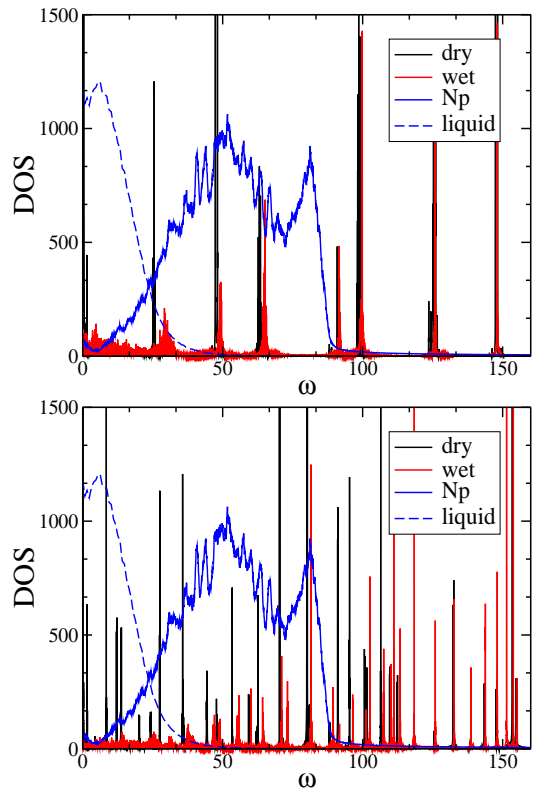

\includegraphics[width=0.8\linewidth,height=0.6\linewidth]{free_junction_n5k3000a1000_DOS_art.eps}
\includegraphics[width=0.8\linewidth,height=0.6\linewidth]{free_junction_n20k3000a1000_DOS_art.eps}
\caption{
Vibrational spectra of dry (black lines) and wet (red lines) junctions with
$N=5$ (top) and $N=20$ (bottom), superimposed on the DOS for the NPs (blue
solid lines) and that of the liquid (dashed blue lines).}
\label{DOS_NP_junction_dry_wet}
\end{figure}

We discuss, finally, how the length of the junction affects the energy
transfer between the NPs. The variation of the conductivity of dry and wet
junctions with length is displayed in Fig.~\ref{power_length}. The stiffness
$k$ ($=6000$) has been chosen here so as to ensure a good matching between
the junction and the thermal contacts with the NPs. For the dry junction, the
relation between conductivity and length is monotonous: the conductivity
increases almost linearly, i.e., the conductance is approximately constant.
This behaviour is standard for unidimensional systems, and easy to understand. Upon increasing the length of the
junction, the number of vibrational modes $N_{\rm modes}=2N-3$ increases
proportionally and the overlap with the NP DOS increases, provided  the
normal mode frequencies $\omega_{\rm stretching} = \sqrt{k/m}$ and
$\omega_{\rm bending} = \sqrt{\kappa/l_0^2}$ are not too small. This is
demonstrated in Fig.~\ref{DOS_NP_junction_dry_wet} where we superimpose the
spectra of free dry junctions (not connected to nanoparticles)
of two different lengths ($N=5$ and $N=20$) with the NP DOS: the overlap
clearly increases with length and, as a result, the conductivity  increases.

For wet junctions, the scenario is different, as can be seen in
Fig.~\ref{power_length}. The conductivity first increases with length,
following closely the results for the dry junction; for $N>13$, however, the
current starts decreasing with increasing  length, i.e. the conductance drops. This
non monotonous behaviour may be understood by analysing the vibrational
spectra of wet junctions, also shown in Fig.~\ref{DOS_NP_junction_dry_wet}.
These have been calculated from simulations of a free junction (i.e., without
NPs) immersed in a liquid.
For the shortest junction ($N=5$), the DOS of wet and dry junctions are very
similar, except perhaps at low frequencies; this is evidently consistent with
the fact that the conductivities of short junction is not affected by the
presence of the surrounding solvent, as observed above
(Fig.~\ref{power_length}). Fig.\ \ref{DOS_NP_junction_dry_wet} however shows
that the spectra of long wet and dry junctions may be different and, in
particular, the low-frequency modes of the dry junction are absent in the wet
junction. We infer from this observation that the surrounding solvent damps
the low-frequency modes of the wet junction. To ascertain this, we also show
in Fig.\ \ref{DOS_NP_junction_dry_wet} the DOS of the Lennard-Jones liquid,
here thermostatted at the reference temperature $T_0=0.75$. It is found to be
significant only at low frequencies --- $\omega$ less than 40 or so; in
contrast, for higher frequencies, the spectrum of the liquid is vanishingly
small. Hence, high-frequency modes of the junction feel the presence of
the solvent as a frictionless elastic medium, and are thus not affected by
the presence of the liquid. This explains why the high-frequency part of the
spectrum of dry and wet junctions are very similar, irrespective of the length
of the junction. Low-frequency modes, on the other hand, feel the solvent as
a viscous medium, and are therefore significantly damped by the solvent. This
is can in fact be seen in Fig.~\ref{DOS_NP_junction_dry_wet} where the
low-frequency modes are almost completely absent, more precisely almost
completely damped --- in the spectrum of the long wet junction.

\section{Conclusion}

We have studied the problem of heat transfer accross a molecular junction
between two nanoparticles that lie either in vacuum and in a liquid
environment using classical molecular dynamics simulations. While our model
remains ``generic'', it does capture the essential physics of the problem.
For the junction in vacuum, the well-known result of a conductance that is
independent of length is recovered; however, the conductivity  depends sensitively
on the overlap between the density of states of the junction itself
and that of the nanoparticles to which it is connected. The presence of a liquid
is found to affect this behaviour in a significant manner: the conductance is
constant to a crossover length, above which it starts to decline and the
standard Fourier regime is recovered. Our results can be rationalized in
terms of the overlap between the vibrational spectrum of the junction and
that of the immersing liquid, which is qualitatively different in short and
long junctions.


\begin{acknowledgments}

We acknowledge funding from the \textit{Agence Nationale de la Recherche}
(ANR) project Opthermal, as well as the Natural Sciences and Engineering
Research Council of Canada (NSERC) and the \textit{Fonds Qu\'{e}b\'{e}cois de
la Recherche sur la Nature et les Technologies} (FQRNT). Part of the
simulations have been done at the P\^ole Scientifique de Mod\'elisation Num\'erique
(Lyon) using the LAMMPS package freely available at http://lammps.sandia.gov
--- see \cite{plimpton1995}

\end{acknowledgments}

\end{document}